\documentclass[%
 reprint,
 amsmath,amssymb,
 aps,
floatfix,
]{revtex4-2}

\usepackage{graphicx}
\usepackage{dcolumn}
\usepackage{bm}
\usepackage{listings}
\usepackage[utf8]{inputenc}
\usepackage{float}

\usepackage{url}

\begin{document}

\preprint{APS/123-QED}

\title{Robust Folding of Elastic Origami}

\author{M. E. Lee-Trimble}
\affiliation{%
 Department of Physics, University of Massachusetts Amherst, Amherst, MA, 01003
}%

\author{Ji-Hwan Kang}
\affiliation{
 Department of Chemical Engineering, California State University Long Beach, Long Beach, CA, 90840\\
 Department of Polymer Science and Engineering, University of Massachusetts Amherst, Amherst, MA, 01003
}%

\author{Ryan C. Hayward}
\affiliation{%
 Department of Chemical and Biological Engineering, University of Colorado Boulder, Boulder, CO, 80309\\
 Department of Polymer Science and Engineering, University of Massachusetts Amherst, Amherst, MA, 01003
}%
\email{ryan.hayward@colorado.edu}

\author{Christian D. Santangelo}
\affiliation{%
 Department of Physics, Syracuse University, Syracuse, NY, 13244\\
 Department of Physics, University of Massachusetts Amherst, Amherst, MA, 01003
}%
\email{cdsantan@syr.edu}

\date{\today}

\begin{abstract}
Self-folding origami, structures that are engineered flat to fold into targeted, three-dimensional shapes, have many potential engineering applications. Though significant effort in recent years has been devoted to designing fold patterns that can achieve a variety of target shapes, recent work has also made clear that many origami structures exhibit multiple folding pathways, with a proliferation of geometric folding pathways as the origami structure becomes complex. The competition between these pathways can lead to structures that are programmed for one shape, yet fold incorrectly. To disentangle the features that lead to misfolding, we introduce a model of self-folding origami that accounts for the finite stretching rigidity of the origami faces and allows the computation of energy landscapes that lead to misfolding. We find that, in addition to the geometrical features of the origami, the finite elasticity of the nearly-flat origami configurations regulates the proliferation of potential misfolded states through a series of saddle-node bifurcations. We apply our model to one of the most common origami motifs, the symmetric ``bird's foot,'' a single vertex with four folds. We show that though even a small error in programmed fold angles induces metastability in rigid origami, elasticity allows one to tune resilience to misfolding. In a more complex design, the ``Randlett flapping bird,'' which has thousands of potential competing states, we further show that the number of actual observed minima is strongly determined by the structure's elasticity. In general, we show that elastic origami with both stiffer folds and stiffer faces self-folds better.
\end{abstract}

\maketitle


\raggedbottom
\section{Introduction}
Origami-like self-actuating structures are found throughout nature \cite{burgert_actuation_2009, saito_investigation_2017} and have inspired a number of engineering applications from medicine \cite{kuribayashi-shigetomi_cell_2012, kuribayashi_self-deployable_2006}, to solar panel deployment \cite{miura_method_1985}, to robotics \cite{kotikian_untethered_2019, onal_origami-inspired_2013}. It has become clear, however, that the space of configurations accessible to a rigid origami structure becomes increasingly complicated as the fold pattern itself becomes more complex \cite{chen_branches_2018, stern_complexity_2017}. Additionally, face bending allows access to configurations that would otherwise be impossible in purely rigid origami \cite{schenk2011origami, schenk_geometry_2013, wei_geometric_2013, dudte_programming_2016, filipov_bar_2017,pinson-self-folding}. These features lead to multistability through the proliferation of local minima of the energy \cite{hanna_waterbomb_2014, waitukaitis_origami_2015, silverberg_origami_2015}, sometimes resulting in origami that does not fold easily or repeatably into the target shape, impacting device performance. Various methods to avoid misfolding have been introduced, including biasing the vertices \cite{kang_enabling_2019} or fine tuning individual fold stiffnesses \cite{stern_shaping_2018,tachi_self-foldability_2017}, but the actual mechanisms behind misfolding are still not well understood.

Tachi and Hull have proposed a method to prevent misfolding that takes advantage of the branched structure of the origami configuration space \cite{tachi_self-foldability_2017}. They assume each fold is a torsional spring and adjust the torques induced by the springs to force the origami in a direction in configuration space perpendicular to all undesirable folding pathways. Unfortunately, due to the high dimensionality of the configuration space, there is often no choice of torques that satisfies all of these requirements \cite{chen_branches_2018}.
On the other hand, Stern \textit{et al.}  explored a large class of origami structures made from only quadrilateral faces and, even in this restricted set, found a proliferation of energy mimima \cite{stern_complexity_2017,stern_shaping_2018}. However, most rigid quadrilateral origami cannot be folded at all \cite{dudte_programming_2016}, and so these energy minima represent configurations involving stretching rather than distinguishing between several valid branches. This suggests a more careful treatment of elasticity in origami is crucial to uncovering the mechanisms of misfolding.

In this paper, we compute energy landscapes of weakly folded origami using a bar-and-hinge model of self-folding that includes both face stretching and face bending \cite{schenk2011origami,silverberg_origami_2015, dudte_programming_2016}. Energy landscapes provide a detailed picture of the vicinity of the flat state, where multiple origami branches meet. We show that the mechanisms governing the formation of competing local energy minima are poorly captured by the assumption of rigid, unstretchable origami. Instead, the undesirable energy minima that compete with the target configuration are regulated by saddle-node bifurcations nucleated near the unfolded state, even when the target configuration is very folded. Our model allows us to determine how the ``foldability'' of an origami design is determined both by the stretching and bending moduli of the faces: more bendable faces allow additional folding pathways while more stretchable faces induce saddle-node bifurcations that reduce the number of local energy minima. Critically, the reduction in the number of energy minima does not arise from transitions between branches that induce large strains in the faces but is, nevertheless, enabled by small amounts of strain while the origami is barely folded. Our analysis leads to new insights on the robustness of the target folding pathway to programming errors in the target fold angles.

To go beyond our theoretical analysis, we also demonstrate these effects experimentally on self-folding origami structures using a previously-published trilayer swelling gel system  \cite{na_programming_2015, kang_enabling_2019}. These experiments demonstrate the possibility of using face stiffness to tune the metastability of self-folding origami. 
In contrast, the few methods that have been proposed to prevent misfolding require a more careful tuning of fold angles and stiffnesses \cite{tachi_self-foldability_2017,stern_complexity_2017,stern_shaping_2018}. We demonstrate that robust folding can still be induced in systems where such precise control may not be possible.

\section{Theory of self-folding origami}
\subsection{Folding rigid origami}
Rigid origami, having both unbendable and unstretchable faces, can be modeled as a triangulated surface with $V$ vertices joined by $N$ edges of fixed length spanned by $F$ polygonal faces with additional torsional springs. Because each face is decomposed into triangles, the length constraints of the edges also preserve the sector angles of the faces. We assume there are $N_B$ edges that are adjacent to a single face which we dub boundary edges to distinguish them from the $N_F$ edges that adjoin a pair of faces, which we refer to as folds. Many origami fold patterns do not have triangular faces, however. In those cases, we decompose each face into triangular subfaces along their shortest diagonals \cite{schenk_geometry_2013, wei_geometric_2013, dudte_programming_2016}. This suggests a further division of folds into ``face folds'', those folds spanning a rigid face, and ``active folds'', which drive the self-folding of the origami.

Since the faces are triangular, the state of any origami structure can be represented completely by its fold angles, $(\rho_1, \cdots, \rho_{N_F})$, where each angle $\rho_i$ is the supplement of the corresponding dihedral angle made by the faces adjacent to the edge. We introduce self-folding by incorporating torsional springs on the folds of the form
\begin{equation}\label{eq:bending}
    E_B = \frac{1}{2} \sum_{I\le N_F} \kappa_{B,I} (\rho_I - \bar{\rho}_I)^2,
\end{equation}
where $\kappa_{B,I}$ is the torsional modulus of the $I^{th}$ fold and $\bar{\rho}_I$ the equilibrium angle of the fold. For face folds, we require $\bar{\rho}_I = 0$ so that Eq. (\ref{eq:bending}) penalizes bending of the faces. On active folds, however, $\bar{\rho}_I \ne 0$ which imposes a bending torque that drives the origami to fold along its active folds.

For models of this type, there are singular configurations where several branches of allowed configurations meet \cite{tachi_self-foldability_2017, chen_branches_2018, berry_topological_2020}. Each branch has a tangent space where it meets the singular configuration and many such branches meet at this point \cite{chen_branches_2018}. Tachi and Hull have proposed that misfolding can be prevented when the torque, defined by $\tau_I = - \kappa_{B,I} \bar{\rho}_I$, is perpendicular to the tangent space of each branch \cite{tachi_self-foldability_2017} in the space of folds. Indeed, examining Eq. (\ref{eq:bending}) shows that
the Tachi-Hull condition is precisely the condition that there is no direction along an undesirable branch along which the energy decreases.

It is notable that the Tachi-Hull condition can be impossible to satisfy if the origami fold pattern is sufficiently complicated, as the number of branches grows exponentially with the number of vertices while the number of folds grows linearly \cite{chen_branches_2018}. If there are several branches along which $E_B$ decreases, each of these branches must have at least one local energy minimum. Consequently, the number of potential competing energy minima in a rigid, self-folding origami system can be quite sensitive to even small errors in the programmed torques $\tau_I$. The question of stability and metastability of an origami folding becomes even more complex when one considers Eq. (\ref{eq:bending}) along an entire origami trajectory, and such an analysis has only been undertaken for some single origami vertices \cite{waitukaitis_origami_2015}.

\subsection{Elastic origami}
Part of the sensitivity of competing minima to torques arises from the singular nature of the unfolded, flat origami. To study this further, we augment our model to allow for stretching. We supplement Eq. (\ref{eq:bending}) with additional terms \cite{schenk2011origami} of the form,
\begin{equation}\label{eq:stretching}
    E_S = \frac{1}{2} \sum_{i \le N} \kappa_{S,i} \gamma_i^2,
\end{equation}
where $K_{S,i}$ the stiffness of edge $i$ and the elastic strain is given by
\begin{equation}
    \gamma_i=\frac{1}{2}\left(\frac{L_i^2}{\Bar{L_i}^2}-1\right).
\end{equation}
Note that a small deformations of the edges $\Delta \ll \bar{L}_i$ gives $\gamma_i \approx \Delta/\bar{L}_i$ as does the slightly more common form for the elastic strain $\gamma_i = L_i/\bar{L}_i - 1$.
By formulating the energy in terms of a dimensionless strain, $\kappa_{S,i}$ has the same units as $\kappa_{B,I}$ evaluated on the same edge ($i=I$). We set $\kappa_{S,I} = Y_{2D} \bar{A}_I$ where $Y_{2D}$ is the two dimensional Young's modulus of the origami faces and $\bar{A}_I$ is a characteristic face area. Here, we will set $\bar{A}_I$ to one third the total area of the faces adjoining edge $I$, which implies that $A_I \propto L_I$, and that edges can be subdivided without changing the energy cost of a given strain.
There are more complex choices for $\kappa_{S,I}$ that are expected to capture more detailed features of the stretching deformations \cite{filipov_bar_2017}. As an alternative, we also consider a more realistic model in which the faces themselves are elastic polygons that deform affinely, finding good agreement with our simpler model (see SI). The advantage of Eq. (\ref{eq:stretching}) is that it allows us to make contact with the rigidity theory of frameworks on which the analysis of branched configuration spaces has been done \cite{chen_branches_2018, waitukaitis_origami_2015}.

Eq. (\ref{eq:stretching}) also provides a convenient geometrical interpretation of the  stretching energy of weakly-folded origami in terms of the Gaussian curvature of the vertices \cite{berry_topological_2020}. In the limit that $\kappa_{S,I} \gg \kappa_{B,I}$, we obtain an approximate expression for the energy valid when the fold angles are small. To do so, we note that vertical motions of the vertices off the $xy-$plane preserve $L_I$ to lowest order. Therefore, we can express $E_S$ as a function of the $V_B + V_I - 3$ vertex heights only, where $V_B$ is the number of vertices adjoining a boundary edge and $V_I$ are the number of vertices adjoining only folds. This expression for the energy, quartic in the vertex heights and can be expressed as
\begin{equation}\label{eq:vertexstretching}
    E_S = \frac{1}{8} \sum_{n \le V_I} \left( \mathbf{h}^T \mathbf{Q}_n \mathbf{h}\right)^2
\end{equation}
for a vector of vertex heights $\mathbf{h}=(h_1,...,h_{V_I})$ where $h_i$ is the height of the $i^{th}$ vertex and $\mathbf{Q}_n$ a symmetric matrix which encodes the geometrical constraints associated with the branches as well as the stiffnesses of the origami (see SI for details). One can show that $E_S = 0$ if and only if the discrete Gaussian curvature of each origami vertex vanishes, and that the matrices $\mathbf{Q}_n$ have two zero eigenvalues, one eigenvalue of either positive or negative sign, and the rest of the opposite sign \cite{chen_branches_2018}.

Finally, we consider how to set the relative magnitudes of the torsional spring moduli in our energy. As a model of self-folding origami, we consider a trilayer polymer system described previously \cite{na_programming_2015}, in which faces are characterized by a hydrogel of thickness $h_N$ sandwiched between two stiffer layers $h_P \ll h_N$ and active folds are induced by cutting trenches in either the top or bottom of the two stiff layers of a given width. To estimate the bending rigidity for the faces, we imagine that the bending energy arises from bending along a cylinder oriented along each fold of characteristic width $W_I$. Then face folds will have $\bar{\rho}_I = 0$ and torsional moduli $\kappa_{face,I} \approx Y_p h_N^2 h_P \bar{L}_I/(W_I (1-\nu^2))$ where $Y_p$ is the elastic moduli of the stiff layers, $L_I$ and $W_I$ are the length and width of the fold, and $\nu$ is the Poisson ratio. A similar calculation shows that an active fold has an approximate torsional modulus $\kappa_{fold,I} \approx Y_N h_N^3 \bar{L}_I/(3 W (1-\nu^2))$ where $Y_N$ is the Young's modulus of the hydrogel layer. For the rest of the paper, we will neglect the small changes in $W_I$ between different folds, letting us combine the material parameters and define $\kappa_{face,I} \approx K_{face} \bar{L}_I$ and $\kappa_{fold,I} \approx K_{fold} \bar{L}_I$. We will also define $\kappa_{S,I} \approx K_{S} \bar{L}_I$. The details of these estimates can be found in the Supplementary Information.


\begin{figure}[th]
\centering
\includegraphics[width=0.99\linewidth]{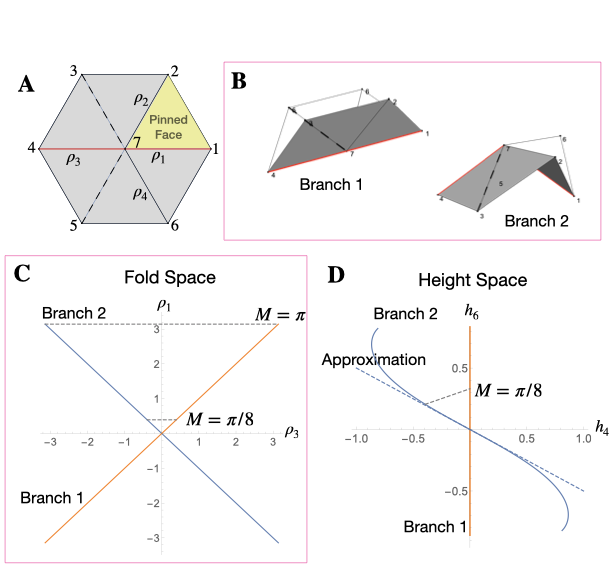}
\caption{Schema, configuration spaces, and energy landscapes for the birdsfoot origami. (A) A schematic of the birdsfoot. The folds are marked in solid lines, while the folds added in the model to imitate face bending are marked with dashed lines. The folds used to define fold space are highlighted in red. The face we have “pinned” to a plane is highlighted in yellow, then the heights of vertices $4$ and $6$ above this plane defines the height space. (B) The configurations that correspond to each branch. (C-D) The branches for a rigid origami in both fold and height space, respectively. The dashed lines show the linearized trajectories between the branches at two magnitudes. The height space visualization of the trajectory takes advantage of the linearity between height and fold space at small heights. In (C), note that the branches in fold space are perpendicular. }
\label{fig:birdsfoot}
\end{figure}

\section{The origami ``bird's foot''}

Eq. (\ref{eq:vertexstretching}) provides a means of computing and plotting energy landscapes for weakly-folded origami. We start our study of the folding and mis-folding of elastic origami with the simplest non-trivial example, the self-folding ``bird's foot'' origami (Fig. 1A). The bird's foot is a single origami vertex from which four folds emerge. We supplement these four folds with two additional face folds, shown as dashed lines in Fig. 1A. It is well known that there are two folding pathways possible which can be characterized by the relative signs of the fold angles between vertices 4 and 1, $\rho_1$ and $\rho_3$ (Fig. 1B).

For the rigid case, in the space of fold angles $(\rho_1, \cdots, \rho_4)$ the trajectories of the fold angles are perpendicular and can be projected conveniently to just a pair of angles as in Fig. 1C. In the elastic case, if we orient the bird's foot so that vertices $1$, $2$ and $7$ per Fig. 1A lie on the $xy-$plane, Eq. (\ref{eq:vertexstretching}) suggests plotting the energy landscape near the unfolded state in terms of the heights of the remaining four vertices above the plane, $\mathbf{h} = (h_3, h_4, h_5, h_6)$ (Fig. 1D) rather than the fold angles.

We program target angles according to
\begin{equation}
    \Vec{\rho}=(1-A) M \Vec{\rho}_{B1}+A M \Vec{\rho}_{B2},
\end{equation}
where $\vec{\rho}_{B1} = ( -1, 0, -1, 0)$ and $\vec{\rho}_{B2} = ( -1, 1, 1, 1 )$ are the fold angles of each branch when folded flat, and $M$ ranges from 0 to $\pi$ and controls the degree of folding. The parameter $A$, which lies between $0$ and $1$, tunes the target angles between the two branches accessible by rigid origami. For values of $A$ other than $0$ and $1$, the target angles lie between the two branches. It should be noted that $A=0.5$ is not precisely between the two branches geometrically, and the geometric center, though dependent on the precise value of $M$, is closer to $A \approx 0.425$. Fig. 1D shows this trajectory in height space for $M=\pi/8$ by taking advantage of the linear relationship between folds and heights to quadratic order \cite{chen_branches_2018}.

\begin{figure*}[th]
\centering
\includegraphics[width=0.99\linewidth]{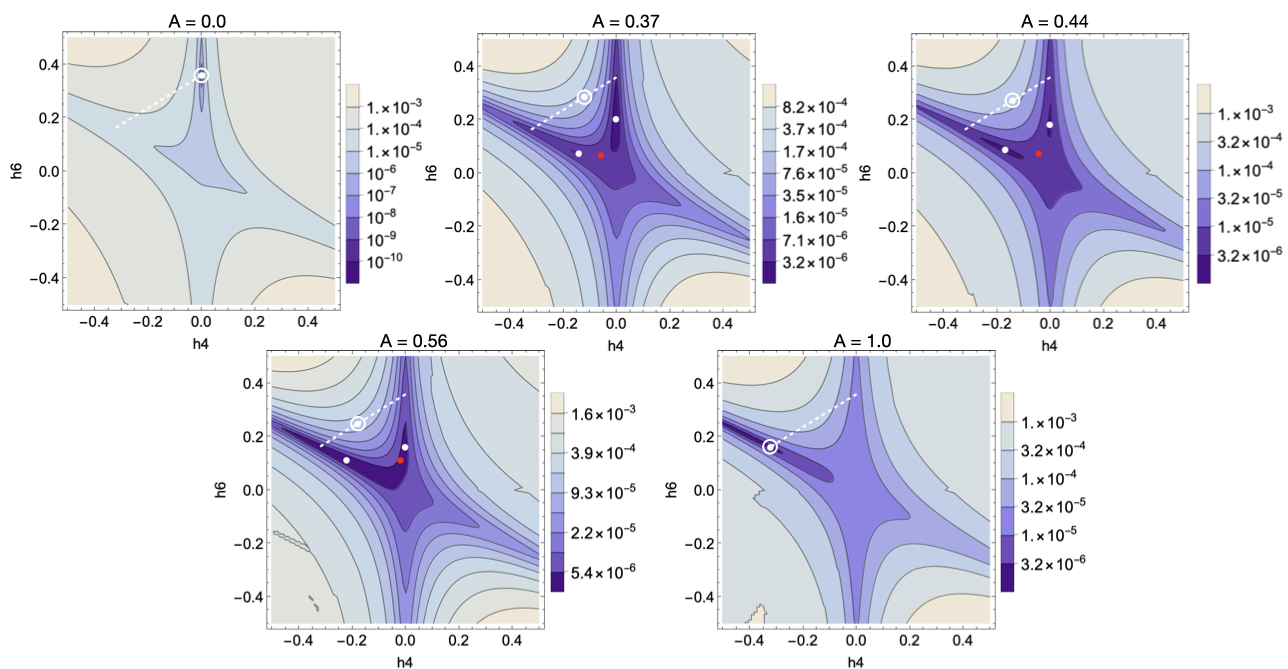}
\caption{Energy landscapes in height space for the bird's foot at different values of the linearization parameter A, for the small magnitude trajectory shown in Fig. 1(D), at $K_{face}=10^{-2}$ and $K_{fold}=10^{-4}$. The white dots represent local minima and the red dots represent saddle points. The dashed line represents the trajectory from Fig. 1(D) with the circle denoting the location on the trajectory of the landscape. Notice that as A increases, the original minimum moves toward the flat state. Between $A=0$ and $A=0.37$, a saddle point and the minimum for branch 2 are created, then after $A=0.56$ the saddle point and the minimum for branch 1 annihilate each other. The created minimum also moves out away from the flat state as A increases. Note that the contours and color scheme are on a log scale and inconsistent between landscapes to emphasize features. }
\label{fig:ori}
\end{figure*}

Since $h_3$ and $h_5$ are the heights of vertices associated with face folds, in order to plot the energy landscapes, we numerically minimize $E(h_3, h_4, h_5, h_6)$ with respect to $h_3$ and $h_5$ to express the energy in terms of only $(h_4, h_6)$. Contours of the energy obtained this way are shown in Fig. 2 for various values of $A$ and for $M = \pi/8$. The minima of the energy are depicted as closed white circles and saddle points are shown in red, while the target point is denoted by an open white circle.

All of the theoretical figures in this paper were created using a package developed for creating and manipulating origami structures and other similar mechanisms in Mathematica. This package is located on GitHub at https://github.com/cdsantan/mechanisms. Mathematica notebooks for each figure and the associated data are also located on GitHub at https://github.com/meleetrimble/robust-folding-paper-support.

As seen in Fig. 2, which shows the energy landscapes of the birds foot with contours on a log-scale, the configuration space of the rigid origami lies along the bottom of steep valleys defined by Eq. (\ref{eq:vertexstretching}). Because the torsional springs are weaker than the stretching springs, as $A$ changes from $0$ to $1$ at fixed $M = 1/8$, the energy minimum on the first branch moves inward along the energy valley. At a critical value of $A > 0$, a new minimum and saddle point nucleate near the flat state and as the new minimum moves outward along the other branch, the old minimum eventually approaches and annihilates with the saddle point.

As the stretching energy is increased, the energy valleys become steeper but the shape of the energy landscape near the flat state remains the same. As $K_S$ increases and we approach the rigid limit, the critical $A$ at which a new minimum forms decreases. Yet for any finite value of $K_S$, the energy landscape is monostable near $A \approx 0$ and $A \approx 1$.

\begin{figure}[t]
\centering
\includegraphics[width=0.99\linewidth]{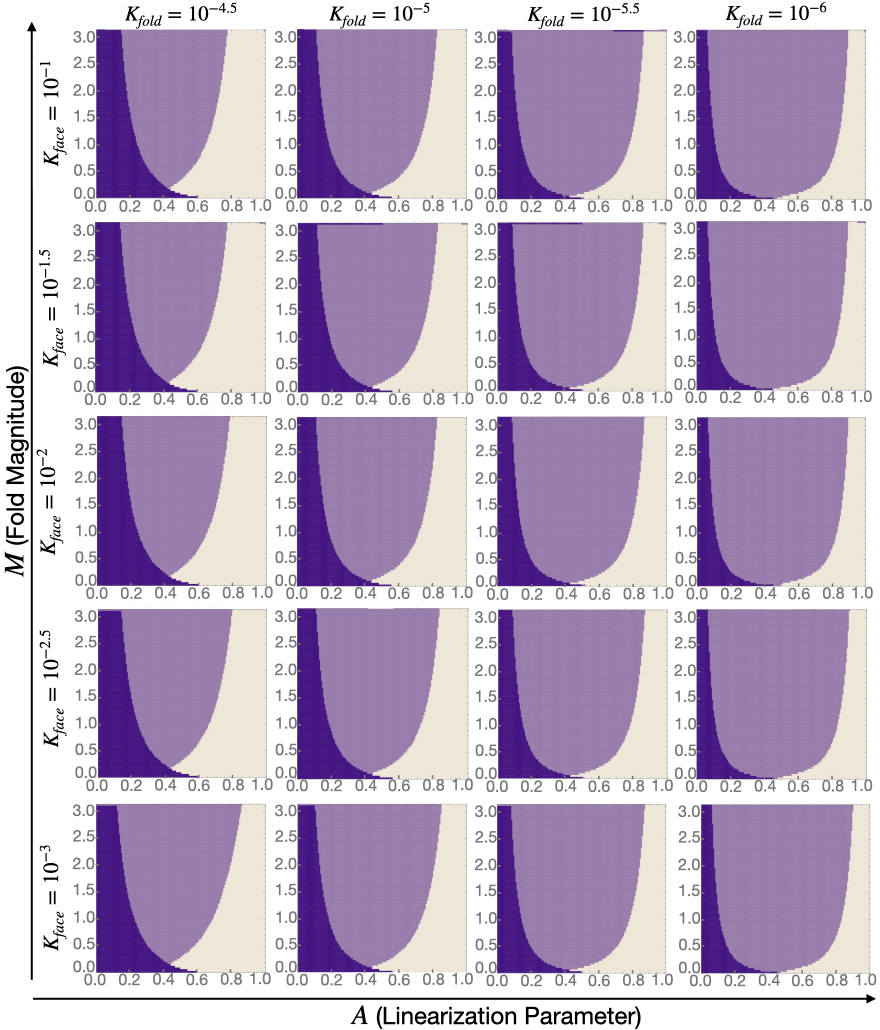}
\caption{Slices of a four-dimensional phase diagram defined by $K_{fold}$, $K_{face}$, linearization parameter $A$, and target fold angle magnitude $M$ at four different values of $K_{face}$ and $K_{fold}$. The regions in light purple represent the region of bi-stability, where both minima are present. The purple and tan represent the regions where only the branch 1 minimum and branch 2 minimum are present, respectively.}
\label{fig:bistab}
\end{figure}

\subsection{Phase Diagrams}

We can determine the size of the region of bistability for different values of $K_{face}$, $K_{fold}$, and prescribed fold angle using the full elastic energy. To do so, we start at one end of the linearization we have introduced and find the minimum at that point. Then we increase the linearization parameter $A$ by one step, and repeat the minimization using the minimum just found as the initial position. We continue taking the next step in the parameterization, using the previous minimum, and minimizing until the other branch has been reached. To see where both minima are present, we repeat this process but instead start from the opposite branch and follow the parameterization backwards. The regime in which both minima are present across both directions is the bi-stable regime.

To get the full idea of the bistable regime between the two branches, we repeat the method described above for different values of $M$, which represents the magnitude of folding, to draw the phase diagram between the two branches as a function of $M$ and $A$ for given elastic moduli.

In Fig. 3, we show the region of bistability as a function of $M$ and $A$ for four different values of $K_{fold}$ and $K_{face}$. The plots are asymmetric and, in particular, shifted toward values of $A<0.5$. This is consistent with the midpoint between both branches being at $A \approx 0.425$ rather than $A=0.5$. Overall, we see two separate trends: decreasing $K_{fold}$ widens the region of bistability with more widening seen at lower values of $M$, while more surprisingly decreasing $K_{face}$ also widens the region of bistability but with more growth seen at higher values of $M$.

These results implicate the balance of in-plane stretching and torsional spring moduli in governing bistability. In particular, when $K_{fold}$ is small, indicating that the system is approaching the inextensible limit, we see that even a small error in programming the fold angles can lead to multistability. This is, in fact, entirely consistent with Ref. \cite{tachi_self-foldability_2017} which argues that for rigid origami, for which $K_{fold}/K_S \rightarrow 0$, a metastable minimum exists unless the vector with components $\kappa_{B,I} \bar{\rho}_I$ is perpendicular to a branch.

It is important to note that the change in bistability occurs even though $K_{fold} \ll K_S$, indicating that in-plane strains are still small and Eq. (\ref{eq:vertexstretching}) remains a reasonable approximation. Indeed, in our simulations the energy from stretching is typically $1 \%$ of the total energy. This is also consistent with the energy landscapes in Fig. 2, which show that the bistability arises from the nucleation of additional minima near the flat state and not far out along a branch even when $M$ is large, precisely where we expect our theoretical analysis of elastic origami to be most accurate.

\subsection{Experimental Methods}
We next turn to a discussion of self-folding in a trilayer, thermo-responsive system \cite{na_programming_2015}, adapted from our previous report \cite{kang_enabling_2019}. In brief, self-folding origami was prepared by using a bilayer bending mechanism of polymer films. P(\textit{p}MS-BP-RhB) (poly(\textit{p}-methylstyrene-benzophenone-rhodamine B) and P(DEAM-AA-BP-RhB) (poly(diethylacrylamide-acrylic acid-benzophenon-rhodamine B) were used as a stiff layer and a thermosensitive hydrogel layer with a lower critical solution temperature at around $30$\textdegree C, respectively. Pendant groups of benzophenone contained in both pre-synthesized co-polymers were utilized as a photoreactive cross-linker for multi-layer patterning. First, the bottom stiff layer was deposited by spin coating of toluene solution of P(pMS-BP-RhB) on a silicon wafer with a water-soluble sacrificial layer of poly(vinyl alcohol) (PVA, Aldrich). 

To create a microscale crease pattern, UV-light (365 nm, pE-100, CoolLED) was projected on the layer of P(pMS-BP-RhB) by an inverted optical microscope (Nikon Eclipse Ti, 10x objective lens) equipped with a digital micromirror device (DMD). Pixelated UV illumination for each layer of birdsfoot pattern was obtained by the Mathematica notebook provided from Robert J. Lang (Tessellatica 11.1d7)\cite{tessellatica} based on the folding angle calibration at a fixed temperature, $20$\textdegree C. After cured, a typical development process was followed by stripping uncured area of the film with a marginal solvent (e.g., mixture of toluene and hexane with 1:3 vol\%). Next, a few-micron thick hydrogel layer was deposited on the sample pattern by casting a chlorobenzene polymer solution and slowly drying in the dark chamber. Patterned UV curing with computer-controlled alignment was then followed for crosslinking of the mid layer on top of the bottom layer. Finally, another thin layer of P(\textit{p}MS-BP-RhB) was photo-patterned as a top stiff layer by using the same procedure as the bottom layer. For folds with a target angle of zero, a series of square holes was additionally applied to all three layers as a perforated line between the vertices of the crease pattern, as shown in Fig. 4A. Because the perforations align on both sides, the resulting face folds have a target angle of $0$ but are stiffer than the active folds represented by slits.

The resultant trilayer origami was fully dried before further use. To release the origami as flat from the substrate, the sample was dipped in the pre-heated buffer solution (pH 7.0 PBS, $60$\textdegree C). After full dissolution of PVA layer, the water bath was cooled down to induce programmed folding of the crease patterns, which was observed by using the optical microscope (Zeiss AxioTech Vario, with 2.5x objective lens).

\subsection{Experimental results}
We can now have a small amount of control over $K_{face}$ by utilizing the perforated $0$ angle folds explained above. Perforating the faces decreases the amount of stiff trilayer by a factor of 3 to 4, and since the stiff layer provides the majority of the bending modulus we expect $K_{fold}/K_{face}$ to decrease by the same factor.

We created batches of $10$ bird's foot origami both with and without perforated faces for several values of $A$, corresponding to different target fold angles, between the two branches. Fig. 4B shows the fraction of bird's foot samples that folded to branch $\vec{\rho}_{B2}$ with non-perforated samples (circles) and perforated samples (squares). In the non-perforated samples, we see a sharp transition between branch 1 at small $A$ and branch 2 at large $A$, with a small region of values near $A \approx 0.5$ that show some bistability. In the perforated samples we see this bistable region widen, with both states observed in the $A \approx 0.33$ samples.

\begin{figure}[t]
\centering
\includegraphics[width=.9\linewidth]{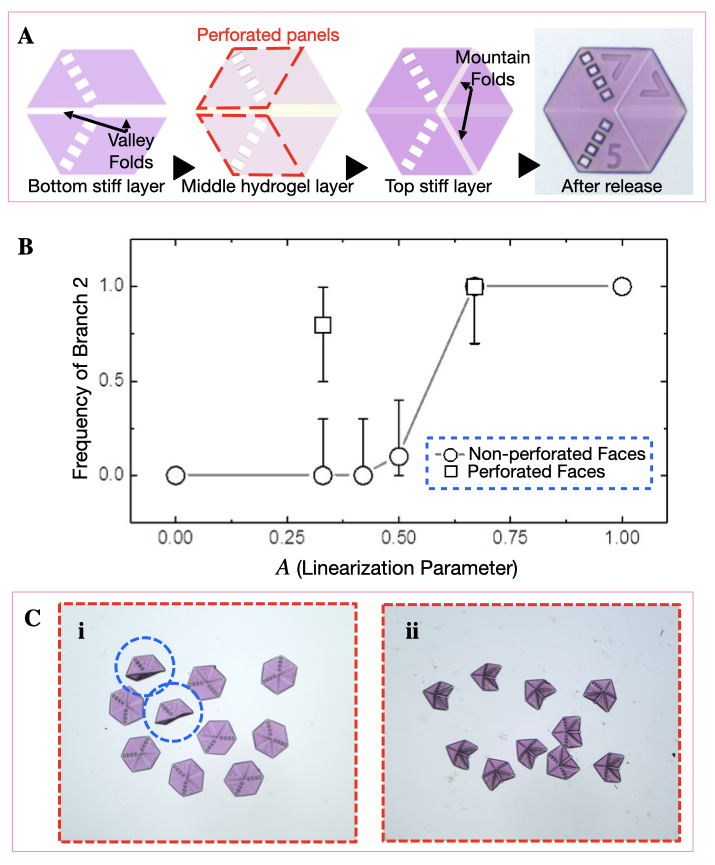}
\caption{Experimental schema and results for folding the birdsfoot with and without weakened faces. (A) Schematic of the tri-layer origami structure with perforated faces. (B) The percentage of samples folded to the second branch for both non-perforated and perforated faces. Each point corresponds to $10$ samples. Error bars are from the rule of three. (C) The folded samples with perforated faces at linearization parameters (i) $A=0.33$ and (ii) $A=0.67$. At $A=0.33$ (i), you can see the two samples that folded to branch $1$ while the rest shallowly folded to branch $2$. 
}
\label{fig:exp}
\end{figure}

Some care must be taken in interpreting the results quantitatively. Because the experimental system folds slowly, we expect the number of minima to be governed to some degree by the small $M$ portion of Fig. 3, even when the programmed fold angles are large, since we expect that a structure that has found a stable configuration will tend to remain in that configuration as it folds. In addition, failure to see misfolded states does not indicate that those states do not exist; in contrast, even a small number of misfolds indicates metastability. Finally, the programmed fold angles are controlled by the width of the long cuts in either the top or bottom rigid layers. The cuts that lead to folding then also affect the torsional stiffness of the folds. This effect is negligible for most folds, except for those with zero fold angle (those remaining flat), which must be cut on both top and bottom surfaces. This leads to folds that are weaker than active folds, as is the case at the $A=0.5$ point. This point does, however, highlight that using target angle tuning to avoid misfolding can be more complicated to realize in experiment than in theory.

To compare our experimental and theoretical results, we estimate that, in units of the length of the folds, $K_S \approx \sqrt{3}/3$ on the folds and $K_S \approx \sqrt{3}/6$ on the boundary edges, $K_{face} \approx 2 \times 10^{-3}$ and $K_{fold} \approx 6 \times 10^{-6}$ (see SI), and use the factor stated above when faces are weakened. According to the relevant region of plots in Fig. 3, perforating the faces should widen the bistable region by weakening the face folds and this effect is seen quite prominently in Fig. 4B, as well as the bistable region occurring for smaller $A$. In Fig. 4C, we show a representative batch of $10$ origami structures. It is also notable that the misfolded configurations in 3C are quite shallow, as we expect from our theoretical analysis. Though the experiments are in qualitative agreement with our theoretical model, the effect of softening the torsional moduli of the faces affects the stability of experimental bird foot origami rather dramatically whereas the theory shows more subtle effects. The origin of this discrepancy remains unclear.

\section{Folding complex origami}
Finally, we turn to a more complex fold pattern, the ``Randlett bird'' \cite{randlett_flapping_1969}, (Fig. 5A-B), which we have previously explored with the trilayer, self-folding origami system \cite{na_programming_2015}. Here, we use the same programmed fold angles from Ref. \cite{kang_enabling_2019} (Supplementary Information). We previously reported that self-folding trilayer Randlett birds misfold at a rate of $0.55 \pm 0.15$ \cite{kang_enabling_2019}. Some examples of both correctly and incorrectly folded birds can be seen in Fig. 5C.

Unlike the bird's foot origami, the Randlett bird is not foldable without bending faces. If we introduce face folds across the shortest diagonal of the faces, however, we expect the Randlett bird to have $2048$ branches each with $6$ degrees of freedom (as predicted by formulas in Ref. \cite{chen_branches_2018}). The high dimensionality of this enlarged configuration space makes direct visualization of the energy landscape impossible. Instead, we will apply a statistical analysis to the folded minima.

We first initialize the Randlett bird in the folded configuration according to vertices provided by Ref. \cite{origamisimulator} (described in \cite{ghassaei2018fast}) and attempt to numerically minimize using the BFGS algorithm \cite{nocedal1980updating}. For $K_{face} \geq 10 K_{fold}$, this direct numerical minimization of the pre-folded state fails. To avoid complications when counting minima, we only use values of $K_{face}$ and $K_{fold}$ for which this minimization produces a reliable energy minimum. The gray region in Fig. 5D represents this region. Note that the expected stiffnesses for the trilayer origami system, $K_{face} \approx 10^{-3}$ and $K_{fold} \approx 6 \times 10^{-6}$, are in this region.

\begin{figure}[H]
\centering
\includegraphics[width=0.77\linewidth]{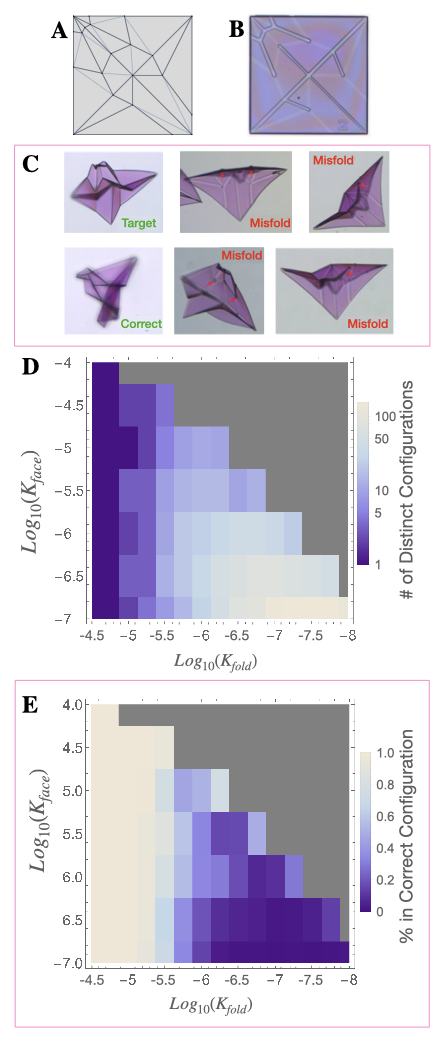}
\caption{Schema, experimental images, and simulation data for the Randlett bird. (A) Schematic of the simulated Randlett bird with added face folds in blue. (B) An optical image of the experiment before folding. (C) Some examples of folded and mis-folded structures. The experiment folds correctly $45 \pm 0.15\%$ of the time. (D) A density plot of the number of different configurations seen from near the flat state for given $K_{fold}/K_S$ and $K_{face}/K_S$. Each point represents a minimum of either $500$ simulated folds or $10$ times the number of different states observed, whichever was larger. The number of configurations reflects whether the origami is mono- or multistable, but may not predict the precise number of possible fold configurations for the given parameters. Note that the color scale is a log scale to emphasize features at lower configuration numbers. The gray region represents the region in which the pre-folded, initial numerical minimization fails. (E) A density plot showing the percentage of the simulated birds that fold into the target state. This does not represent a prediction of experiment, rather that the basin of attraction for the correct state is larger when the degree of multistability is less.}
\label{fig:rand}
\end{figure}

A Mathematica notebook for this subfigure is located on GitHub at https://github.com/meleetrimble/robust-folding-paper-support.

For a given set of $K_{face}$ and $K_{fold}$, we start by generating a sample of 300 randomly perturbed Randlett birds at a given set of $K_{face}$ and $K_{fold}$ by moving each vertex of the bird out of its flat starting position by a normal distribution with a width of $50\%$ of the shortest fold in the origami. We then minimize each bird's energy and discard any results that fail to find a minimum to within a target accuracy goal. We continue to generate further samples until we reach a total of the larger of 500 successful minimizations or $10$ times the number of distinct minima found. We then identify distinct folded states by first determining the optimal alignment by a least-squares minimization of the distance between corresponding vertices of a pair of birds with respect to Euclidean motions, then determining whether all corresponding vertices are closer to each other than a threshold value. This threshold value is chosen so that the number of distinct minima does not change when the threshold value is changed.

Finally, we count the number of distinct states, each representing a mechanically stable state. While there is no way to guarantee that this procedure finds every metastable state, we expect the relative number of energy minima found to scale with the actual number of metastable minima. We then also extract the percentage of samples folded to the target state. We perturbed the simulated samples from the flat state using a normal distribution, so the initial birds represent a uniform cloud of initial states in position space. Thus, this percentage does not represent the folding rate of experiment, but rather the relative size of the basin of attraction for the target minima. 

Fig. 5D shows the resulting number of minima we find as a function of $K_{face}$ and $K_{fold}$ on a log scale to emphasize the points that have only a single minima. Fig. 5E shows the percentage of samples folded to the target minimum for the same data. In both plots, each point represents at least the larger of $500$ birds or $10$ times the number of distinct states seen. The two plots together show that a lesser degree of multistability leads to the basin of attraction for the correct minimum increasing. This implies that there is a relationship between the number of minima and the robustness of the folding origami.

Overall, we see the same effect for the Randlett bird that we saw for the bird's foot: multistability increases with both decreasing $K_{fold}$ and decreasing $K_{face}$. The method to arrive at this result for the Randlett bird can be generalized for any origami, and we would expect the same general result.

\section{Conclusions}
We have introduced a simple model to study self-folding origami that accounts for the finite elasticity of the origami. With finite elasticity, a more complicated picture of the energy landscapes and folding of these structures arises than in rigid origami. Though the energy landscape is characterized by deep valleys along the configuration space of the rigid structures (so that strains while folding are still typically small) we find that the number of energy minima changes with the elastic moduli of the folds through a series of bifurcations near the flat configuration. Because these bifurcations occur near the flat configuration, where finite elasticity dominates the shape of the energy landscape, they are not well-captured by analyses of rigid origami.

We demonstrated two methods for using this model to examine the stability of origami for different stretching and bending parameters: first one that can be applied to simple origami with low-dimensional configuration spaces that can be easily represented, and a second method that can be applied to much more complicated origami. In both cases, we saw that weakening both faces and folds results in an increase in the degree of multistability of the structures. In other words, thicker, elastic origami self-folds better than idealized origami with infinitely stiff faces and floppy folds. 

Both Tachi and Hull\cite{tachi_self-foldability_2017} and Stern \textit{et al.} \cite{stern_shaping_2018} proposed methods for avoiding misfolding that utilize tuning the target fold angles and fold stiffnesses to avoid misfolding. Both methods require a more precise fine-tuning of fold stiffnesses and angles that are often difficult to achieve in many experimental platforms. Tuning the in-plane and out-of-plane stiffnesses of the faces themselves, either by weakening as suggested here or stiffening by adding additional layers, is an additional simple tool to avoid misfolding even when geometric constraints are still dominant. 

\section*{Author Contributions}
CDS and RCH conceived and supervised the research. MEL performed theoretical and numerical analysis. JHK performed experiments. All authors interpreted results.

\begin{acknowledgments}
We thank D. W. Atkinson, I. Griniasty, and N. S. DeNigris for useful discussion. This material is based upon work supported by the National Science Foundation through grant DMR-1822638 and the Graduate Research Fellowship under Grant No. 451512, and by the Army Research Office through grants W911NF-19-1-0348 and W911NF-21-1-0068.
\end{acknowledgments}

\appendix

\section{Elastic moduli of trilayer origami}

Our estimates of elastic moduli will be based on the estimates $Y_N/Y_p \sim 5 \times 10^{-4}$ and $h_p/h_N \sim 0.04$.

\subsection{Estimate of the stretching energy of elastic origami}

We estimate the elastic energy of a face according to
\begin{multline}
E = \frac{1}{2} \int dA \bigg[ \int_{-h_N/2-h_P}^{-h_N/2} dz ~Y_p + \int_{-h_N/2}^{h_N/2} dz~Y_N\\ + \int_{h_N/2}^{h_N/2+h_P} dz~Y_p \bigg] \gamma^2,
\end{multline}
where $Y_N$ and $Y_p$ are the three dimensional Young's moduli and $h_N$ is the thickness of the hydrogel layer,  $h_P$ is the thickness of each polymer layer, and $\gamma$ is a dimensionless strain. 

\begin{figure}[ht]
\begin{center}
\includegraphics[width=0.5 \textwidth]{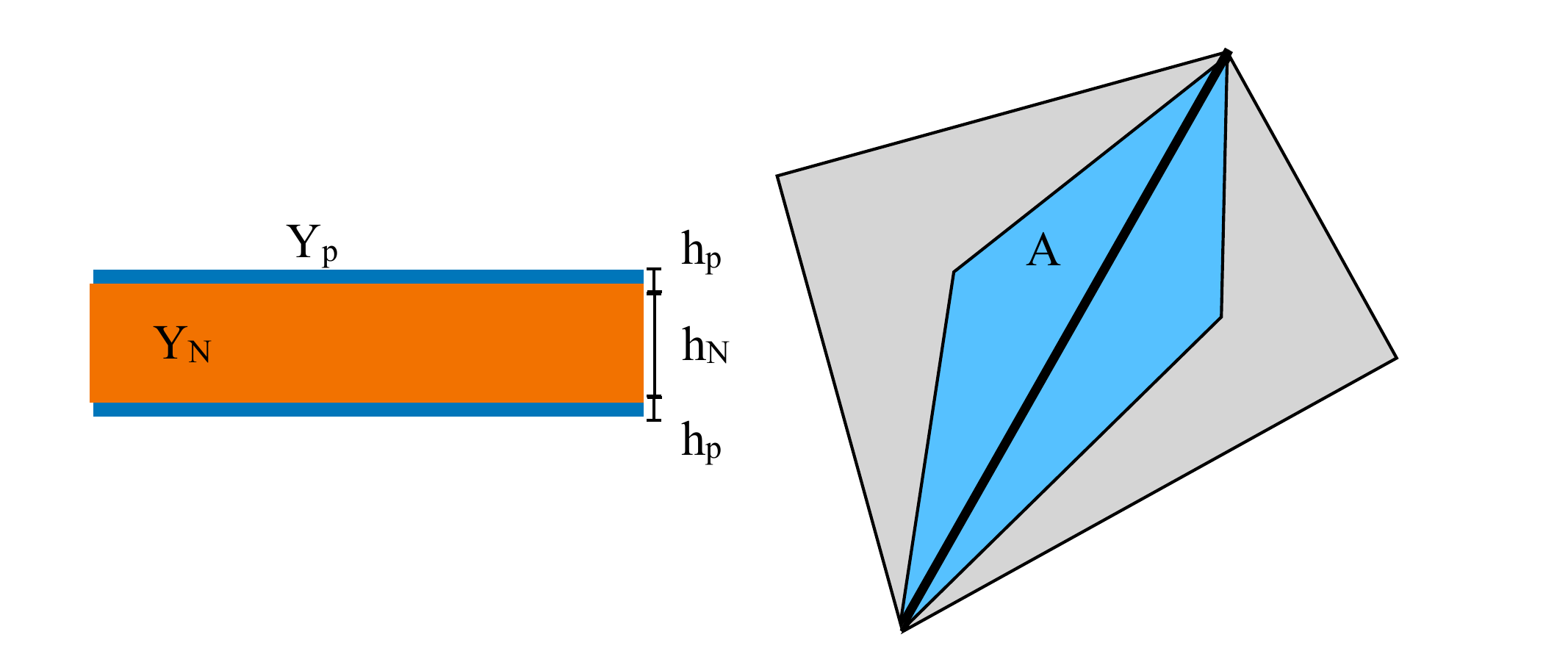}
\caption{(left) a cross-section of a trilayer origami face showing the thicknesses and three-dimensional Young's moduli. (right) the area of each face adjacent to an edge that is closer to that edge than any other.}
\label{fig:stretching}
\end{center}
\end{figure}

Then assuming that $\gamma$ is approximately constant across a face and assuming $Y_p h_P \gg Y_N h_N$, we obtain
\begin{equation}
E \approx Y_p h_{p} A \gamma^2.
\end{equation}
For the area, $A$ and an edge surrounded by two faces, we use the area in Fig. \ref{fig:stretching}, which is conveniently one third the total area of the two adjoining faces. For edges on the boundary, the corresponding stretching energy is obtained from a single face. Comparing this to our spring energy,
\begin{equation}
E = \frac{1}{2} \kappa_S \gamma^2,
\end{equation}
we obtain an estimate $\kappa_{S,I} \approx 2 Y_p h_p A_I$ for the stretching modulus associated with edge $I$, where $A_I$ is the appropriately chosen area.

\subsection{The bending modulus of the folds}

\begin{figure}[ht]
\begin{center}
\includegraphics[width=0.45 \textwidth]{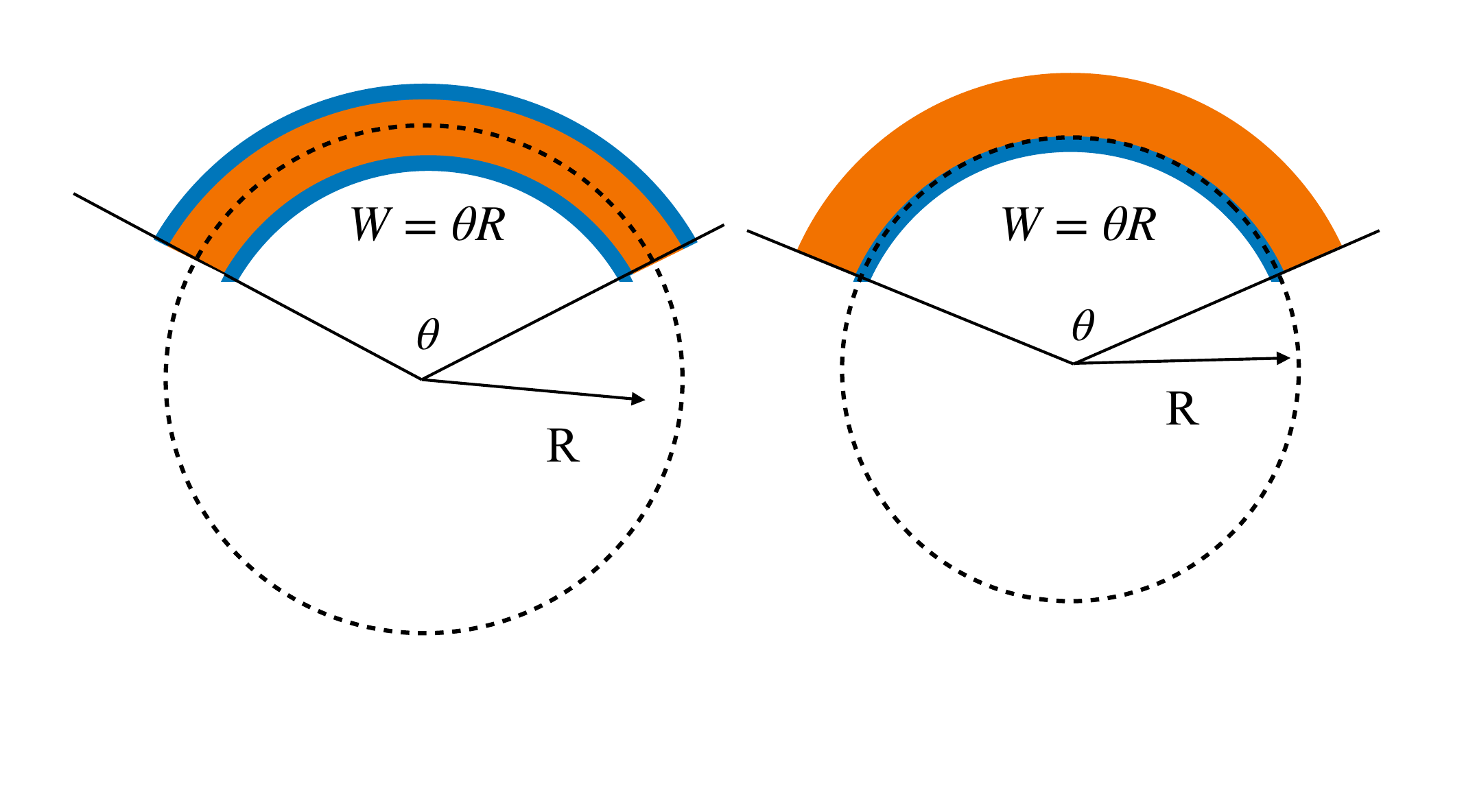}
\caption{Bending a face or a fold to have constant curvature $R^{-1}$. The angle $\theta$ is identical to the apparent fold angle of the fold. When bending a face (left), we assume the face bends along the midsurface whereas for an active fold, we assume the surface bends along the stiffest layer.}
\label{fig:bending}
\end{center}
\end{figure}

For an active fold, we assume the fold is bent along a width $W_I$ to a constant curvature $R$, so that $R = W_I/\theta$. Therefore, the bending energy of a face can be approximately computed as
\begin{multline}
E_B = \frac{1}{2} \frac{Y_N}{1-\nu^2} W L \int_{-h_N/2}^{h_N/2} dz \frac{z^2}{R^2} \\+ \frac{1}{2} \frac{Y_p}{1-\nu^2} W L \int_{-h_N/2-h_P}^{-h_N/2} dz \frac{z^2}{R^2} \\+ \frac{1}{2} \frac{Y_p}{1-\nu^2} W L \int_{h_N/2}^{h_N/2+h_P} dz \frac{z^2}{R^2} \\
\approx \frac{1}{2} \theta^2 L \frac{Y_p h_N^2 h_P}{W (1-\nu^2)}
\end{multline}
Thus, $K_{fold,I} \approx L_I/W_I (Y_p h_N^2 h_P)/((1-\nu^2))$. For an active fold we obtain
\begin{multline}
E_B = \frac{1}{2} \frac{Y_N}{1-\nu^2} W L \int_{h_P/2}^{h_N+h_P/2} dz \frac{z^2}{R^2}\\ + \frac{1}{2} \frac{Y_p}{1-\nu^2} W L \int_{-h_P/2}^{h_p/2} dz \frac{z^2}{R^2}\\
\approx \frac{1}{4} \theta^2 L \frac{Y_N h_N^3}{3 W (1-\nu^2)}
\end{multline}
Thus, $\kappa_{face,I} \approx L_I/W_I Y_N h_N^3/(3 (1-\nu^2))$. We expect that the width of an active fold is set by the size of the cut used to create the folding face whereas the width of a fold associated with bending a face is set by the size of a vertex which is also the width of the trenches. Therefore, we assume $W_I$ is the same for both types of folds.

\subsection{Stiffness ratios}

In our numerical calculations, we divide all the moduli by $2 Y_p h_p A$ where $A$ is the characteristic area. Neglecting the Poisson ratio, we use
\begin{eqnarray}
\kappa_{S,I} &\approx& A_I/A \nonumber \\
\kappa_{fold,I} &\approx& \frac{L_I}{\ell}\left(  \frac{Y_N}{Y_p} \frac{\ell}{W} \frac{h_N^3}{6 h_p A} \right)\\
\kappa_{face,I} &\approx& \frac{L_I}{\ell} \left( \frac{h_N^2 \ell}{2 W A}  \right) \nonumber
\end{eqnarray}
where $\ell$ is the characteristic length of a fold. For $h_P \approx 0.2 \mu m$, $h_N \approx 5 \mu m$, $Y_N/Y_p = 5 \times 10^(-4)$ and $A \approx 260 \mu m$ and $\ell \approx 260 \mu$, we obtain $\kappa_{fold,I} \approx 6 \times 10^{-6} L_I/\ell$ and $\kappa_{face,I} \approx 10^{-3} L_I/\ell$, or $K_{fold} = 6 \times 10^{-6}$ and $K_{face} =  10^{-3}$

\section{Energy of nearly flat origami}
To derive the elastic energy for nearly flat origami in the small strain limit, we assume we have already added face folds so that the origami is built from only triangular faces. We then define a vector function of the vertex positions
\begin{equation}
    f_i(\mathbf{u}) =  \frac{\sqrt{K_i} }{2} \left( \frac{L_i^2}{\bar{L}_i^2} - 1 \right)
\end{equation}
where $\mathbf{u} = (\mathbf{X}_1, \cdots \mathbf{X}_V)$ is a vector containing the position of all $V$ vertices. Then the stretching energy is written as
\begin{equation}
    E_S = \frac{1}{2} \sum_{i=1}^E f_i(\mathbf{u})^2.
\end{equation}

We now expand $f_i(\mathbf{u})$ around the flat state $\mathbf{u}_0$ to find
\begin{multline}
    f_i(\mathbf{u}_0 + \delta \mathbf{u})\\ \approx \partial_n f_i(\mathbf{u}_0) \delta u^n + \frac{1}{2} \partial_n \partial_m f_i(\mathbf{u}_0) \delta u^n \delta u^m
\end{multline}
We next construct an orthonormal basis in the space of possible edges indexed by $i$, $\{ \sigma_{1,i}, \cdots, \sigma_{S, i}, e_{1, i}, \cdots, e_{E-S,i} \}$ where $\sum_i \sigma_{N,i} \partial_i f(\mathbf{u}_0) = 0$. The $\sigma_{N,i}$ are, therefore, the components of the self-stresses of the linkage representing the origami structure.

Now the energy can be written as
\begin{multline}
    E_S = \frac{1}{2} \sum_{N=1}^{E-S} \bigg[ \sum_i e_{N,i} \big( \partial_n f_i(\mathbf{u}_0) \delta u^n\\ + \frac{1}{2} \partial_n \partial_m f_i(\mathbf{u}_0) \delta u^n \delta u^m \big) \bigg]^2\\ + \frac{1}{2} \sum_{N=1}^S \bigg( \sum_i \sigma_{N,i} \frac{1}{2} \partial_n \partial_m f_i(\mathbf{u}_0) \delta u^n \delta u^m \bigg)^2\nonumber
\end{multline}
Finally, we drop higher order contributions to the first term to obtain
\begin{multline}
    E_S = \frac{1}{2} \sum_{N=1}^{E-S} \left( \sum_i e_{N,i}  \partial_n f_i(\mathbf{u}_0) \delta u^n \right)^2 \\+ \frac{1}{8} \sum_{N=1}^S \left( \sum_i \sigma_{N,i} \partial_n \partial_m f_i(\mathbf{u}_0) \delta u^n \delta u^m \right)^2.
\end{multline}
The first term is the harmonic contribution to the energy. For flat origami, we know that these correspond to the in-plane motions. On the other hand, the second term corresponds to the out-of-plane motion of the vertices. Finally, we note that the number of self-stresses $S$ is given by the number of internal vertices $V_I$.

\begin{figure*}[!]
\begin{center}
\includegraphics[width=0.95 \textwidth]{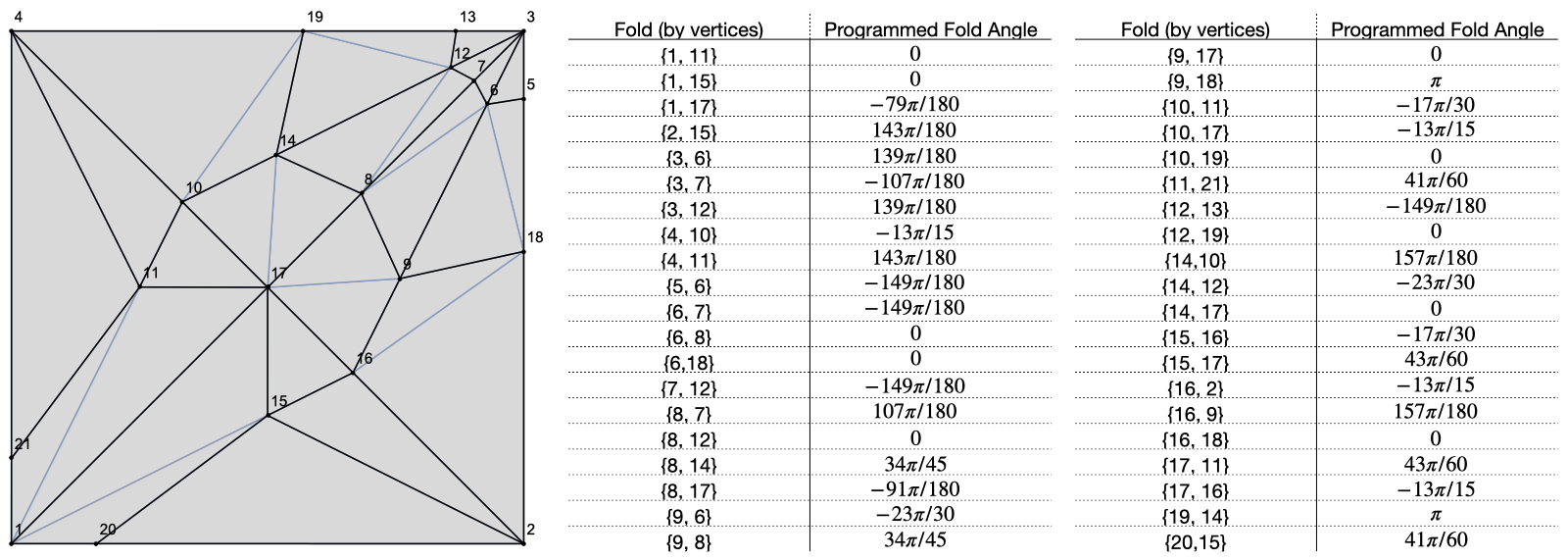}
\caption{(left) the Randlett bird with true folds in black and added face folds in lighter blue with vertices numbered. (right) the programmed fold angles used in the simulations. Folds are denoted by their end vertices.}
\label{fig:randang}
\end{center}
\end{figure*}

For small strains, we assume that the first term is zero so that only out-of-plane deformations can occur. Finally, we obtain an approximate energy for flat origami near the flat state as a sum of terms quartic in the vertical displacements of the vertices,
\begin{equation}
    E_S \approx \frac{1}{8} \sum_{N=1}^{V_I} \left( \sum_{n=1}^V \sum_{m=1}^V Q_{N n m} h_n h_m \right)^2,
\end{equation}
where $h_n$ is the height of the $n^{th}$ vertex above the $xy$-plane and \begin{equation}
    Q_{N n m} = \sum_i \sqrt{K_{S,i}} \sigma_{N, i} \left. \frac{\partial}{\partial h_n} \frac{\partial}{\partial h_m} \gamma_i \right|_{h_n = 0},
\end{equation}
where $\gamma_i$ is the strain of the $i^{th}$ edge defined in the main text.

\section{Alternate model with elastic polygon faces}

The in-plane elastic energy for a Hookean, isotropic two-dimensional solid can be written as
\begin{equation}\label{eq:elasticface}
    E_{el} = \frac{1}{2} \lambda \left(\sum_i \gamma_{ii}\right)^2 + \mu \sum_{ij}\gamma_{ij}^2,
\end{equation}
where $\gamma_{ij} = \partial_i u_j + \partial_j u_i$ and $\lambda$ and $\mu$ are the Lam\`e coefficients. We assume that each triangular face has an energy of the form of Eq. (\ref{eq:elasticface}). The in-plane elastic deformations $u_i$ are determined by assuming the face has deformed affinely. For a triangular face on the $xy-$plane, this uniquely determines the displacement and allows us to estimate the elastic energy of arbitrarily deformed triangular faces.

\section{Programmed target angles for the Randlett bird simulations}

We use the same programmed fold angles for both the experiment and the simulations. They can be seen in Fig. 3.

\vfill

\bibliography{apssamp}

\end{document}